# Structure of nanocrystalline materials using atomic pair distribution function analysis: study of LiMoS$_2$

V. Petkov,[1] S.J.L. Billinge,[1] P. Larson,[1] S.D. Mahanti,[1] T. Vogt,[2] K.K. Rangan[3] and M.G. Kanatzidis[3]

[1] Department of Physics and Astronomy and Center for Fundamental Materials Research, Michigan State University, East Lansing, MI 48823.

[2] Physics Department, Brookhaven National Laboratory, Upton, NY 11973

[3] Department of Chemistry and Center for Fundamental Materials Research, Michigan State University, East Lansing, MI 48823

The structure of LiMoS$_2$ has been experimentally determined for the first time. The approach of atomic pair distribution function (PDF) analysis was used because of the lack of well defined Bragg peaks due to the short structural coherence (∼ 50 Å) in this intercalation compound. This represents one of the first instances where the structure of such poorly diffracting material was determined from the total scattering including diffuse as well as Bragg scattering. The reduction of Mo by Li results in Mo-Mo bonding with the formation of chains of distorted Mo-S$_6$ octahedra. Using the refined structural parameters the electronic band-structure for this material has been calculated and is in good agreement with observed material properties.

Knowledge of the atomic-scale structure is an important prerequisite to understand and predict the physical properties of materials. In the case of crystals it is obtained solely from the positions and intensities of the Bragg peaks in the diffraction data and is given in terms of a small number of atoms placed in a unit cell subjected to symmetry constraints [1]. However, many materials of technological importance are not perfect crystals and have diffraction patterns with a pronounced diffuse component and few Bragg peaks. In this case traditional crystallography fails. For completely disordered materials such as glasses and liquids a statistical description of the structure often is adopted. However, many materials lack perfect long range order and yet have well defined structures on nanometer length-scales (nanocrystals). In this case a description of the atomic structure, in terms of a unit cell and symmetry, can be achieved by employing a non-traditional experimental approach going beyond the Bragg scattering in the diffraction data. An important example is LiMoS$_2$. MoS$_2$ is the key catalyst for the removal of sulfur from crude oil (hydrodesulfurization) [2]. Pristine MoS$_2$ is perfectly crystalline and consists of layers of Mo-S$_6$ trigonal prisms held together by Van der Waals forces. LiMoS$_2$ has Li intercalated between the MoS$_2$ layers. It is important as a precursor of stable MoS$_2$ colloids used to prepare a variety of lamellar nanocomposites [3]. Despite being extensively studied for the last two decades [3-6] the structure of LiMoS$_2$ has not been determined. The reason is that, on Li intercalation, pristine MoS$_2$ is dramatically modified resulting in a product that is too poorly diffracting to allow a traditional structural determination. This leaves unanswered the important question of what exactly happens when MoS$_2$ gets reduced with Li. Theoretical predictions [5] and XAFS [6] studies suggest the presence of considerable Mo-Mo bonding but no unequivocal experimental evidence has been advanced so far. Here we use the atomic pair distribution function (PDF) technique to solve the structure of LiMoS$_2$. This approach has been used widely to study the structure of amorphous materials [7] and to find local structural deviations from a well defined average structure [8]. Here we demonstrate that it also can be used to solve previously unknown structures of poorly diffracting, disordered and nanocrystalline materials. We report for the first time a complete structural determination of LiMoS$_2$.

We find that the structural coherence is limited to ∼ 50 Å in LiMoS$_2$ and, in this sense, the material is nanocrystalline. Nevertheless, it has a well defined local structure with Mo atoms residing in distorted octahedra of sulfur. Short and long Mo-Mo distances appear indicative of metal-metal bonding that is theoretically predicted and easily understood in terms of simple electron counting arguments. Using the experimentally derived atomic coordinates we have calculated the electronic band structure that suggests LiMoS$_2$ is, in fact, a narrow band-gap semiconductor in good agreement with observation.

The atomic PDF is a function that gives the number of atoms in a spherical shell of unit thickness at a distance $r$ from a reference atom. It peaks at characteristic distances separating pairs of atoms and thus reflects the structure of materials. The PDF, $G(r)=4\pi r[\rho(r)-\rho_o]$, is the sine Fourier transform of the so-called total scattering structure function, $S(Q)$,

$$G(r)=(2/\pi)\int_{Q=o}^{Q_{max}} Q[S(Q)-1]\sin(Qr)dQ, \qquad (1)$$

where $\rho_o$ is the average atomic number density, $\rho(r)$ the atomic pair density, $Q$ the magnitude of the scattering vector and $S(Q)$ is the corrected and properly normalized powder diffraction pattern of the material [7]. As can be seen from Eq. 1, $G(r)$ is simply another representation of the diffraction data. However, exploring the experimental data in real space is advantageous, especially in the case of materials with reduced long-range order. First, as Eq. 1 implies the *total scattering,* including Bragg as well as diffuse scattering, contributes to the PDF. In this way all diffracted intensities are considered on the same footing which is critical in the case of a nanocrystalline materials, such as LiMoS$_2$, where the distinction between Bragg and diffuse scattering is not clear. Second, by accessing high values of Q, experimental PDF's of high real-space resolution can be obtained that reveal fine structural features [9]. Third, the PDF is obtained with no assumption of periodicity. Thus materials of various degrees of long-range order, ranging from perfect crystals to glasses and liquids, can be studied using the same approach. Finally, the PDF is a sensitive structure-dependent quantity giving directly the relative positions of atoms in materials. This, as demonstrated in the present Letter, enables the convenient testing and refinement of structural models.

Two samples were investigated. One was pristine MoS$_2$ purchased from CERAC. The second was LiMoS$_2$ obtained by reacting pristine MoS$_2$ with excess LiBH$_4$ [4]. The two powder samples were carefully packed between Kapton foils to avoid texture formation and subjected to diffraction experiments using x-rays of energy 29.09 keV ($\lambda$=0.4257 Å). The measurements were carried out in symmetric transmission geometry at the beamline X7A of the National Synchrotron Light Source (NSLS), Brookhaven National Laboratory. Scattered radiation was collected with an intrinsic germanium detector connected to a multichannel analyzer. The raw diffraction data were corrected for flux, background, Compton scattering, and sample absorption. They were then normalized to obtain the structure function *S(Q)* [7]. All data processing was done using the program RAD [10]. Experimental structure functions are shown in Fig. 1 on an absolute scale and their Fourier transforms, the PDFs *G(r),* in Fig. 2.

Sharp Bragg peaks are present in the *S(Q)* of MoS$_2$ up to the maximal *Q* value of 24 Å$^{-1}$ (Fig. 1(b)). The corresponding *G(r)* also features sharp peaks reflecting the presence of well-defined coordination spheres in this "perfectly" crystalline material (Fig. 2(b)). The well-known 6-atom hexagonal unit cell of MoS$_2$ [11] was fit to the experimental PDF and the structure parameters refined so as to obtain the best agreement between the calculated and experimental data. The fit was done with the program PDFFIT [12] and was constrained to have the symmetry of the *P6$_3$/mmc* space group. In comparing with experiment, the model PDF was convoluted with a *Sinc* function to account for the finite *Q$_{max}$* of the measurement. The best fit achieved is shown in Fig. 2(b) and the corresponding value of the goodness-of-fit indicator R$_{wp}$ is 21 %. [13]. The refinement results are summarized in Table I and the agreement with the published crystallographic results [11] is very good.

We turn now to the LiMoS$_2$ data. In this material the

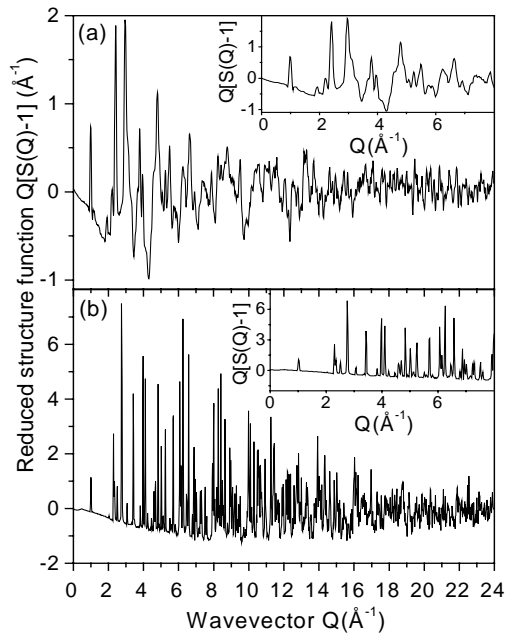

FIG 1. Experimental structure functions of (a) LiMoS$_2$ and (b) MoS$_2$. Note the different scale between (a) and (b). The data are shown in an expanded scale in the insets.

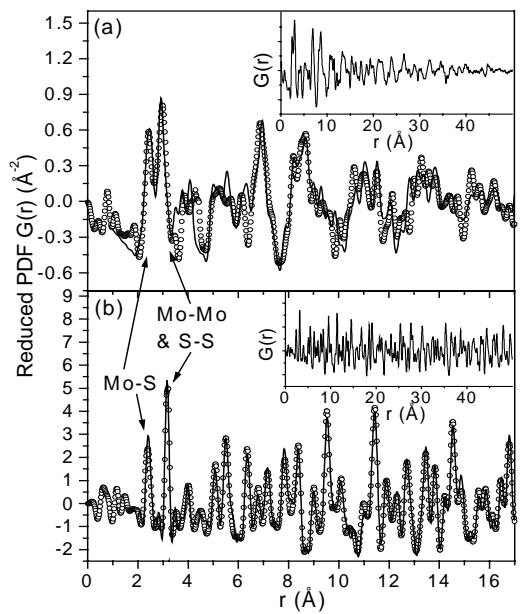

FIG 2. Experimental (dots) and fitted (solid line) PDFs for LiMoS$_2$ (a) and MoS$_2$ (b). The first two peaks in the PDFs are labeled with the corresponding atomic pairs. The experimental data are shown in an expanded scale in the insets.

Bragg peaks are significantly broadened (see the inset in Fig. 1(a)) and already at ~ 8 Å$^{-1}$ merge into a slowly oscillating diffuse component. A diffraction pattern so poor in sharp Bragg peaks lacks the number of "statistically independent reflections" [14] needed to apply the traditional techniques for structure determination. The corresponding PDF is, however, rich in distinct, structure related features and, as we will see, lends itself to structure determination. In the inset in Fig. 2(a) it is evident that the features in the PDF disappear at 50 Å. In this sense, $LiMoS_2$ is nanocrystaline. However, its local structure is still relatively well defined with clearly identifiable Mo-S and Mo-Mo coordination spheres as seen in Fig. 2(a). The structural features persist to much higher r values in the pristine $MoS_2$ (see the inset in Fig. 2(b)) which is a macroscopic crystal.

Table. I. Structural parameters for $MoS_2$. Space group is $P6_3/mmc$. Mo is at (1/3,2/3,1/4) and S at (1/3,2/3,z).

|   | PDF refinement | Single crystal [11] |
|---|---|---|
| a(Å) | 3.169(1) | 3.1604(2) |
| c(Å) | 12.324(1) | 12.295(2) |
| z | 0.623(1) | 0.629(1) |

To determine the parameters of the atomic structure of $LiMoS_2$ we explored several structure models as follows: First, a model based on the hexagonal $MoS_2$ structure (*S.G. $P6_3/mmc$*) was attempted. It was constructed by inserting 2 Li atoms into the 6-atom unit cell of $MoS_2$ so that they resided between the $Mo-S_2$ layers. This model, however, could not reproduce the experimental data as can be seen in Fig. 3(a). A previously proposed [4] structure for $LiMoS_2$ is based on trigonal $1T-MoS_2$ (*S.G. P3*) which can be considered as built of layers of distorted $Mo-S_6$ octahedra. This structure model was also tested with the starting structure parameters, lattice constants and atomic coordinates of Mo and S, being those reported in [6]. Again Li atoms were added to occupy appropriate sites between the $Mo-S_2$ layers. The model performed somewhat better but still could not reproduce the details in the experimental PDF as can be seen in Fig. 3(b). Therefore $LiMoS_2$ and $1T-MoS_2$ are not exactly of the same structure type, contrary to what has been suggested before [4]. Next we attempted a model based on the "exfoliated-restacked" $MS_2$ structure (M= W, Mo) [15] which features zig-zag metal-metal chains. This was an excellent starting point and as the refinement proceeded the Mo atoms shifted slightly but significantly to new positions that define the chains of Mo atoms shown in fig. 4. This model fits all the important details in the experimental PDF as can be seen in Fig. 2(a). The refined structure parameters are summarized in [16]. The agreement with the data is less satisfactory than the $MoS_2$; however this model is significantly better than either of the previously proposed models. Given the high degree of disorder inherent in $LiMoS_2$ this level of agreement is quite good and acceptable.

In pristine $MoS_2$ Mo atoms from a single $MoS_2$ layer arrange in a regular hexagonal lattice (see Fig. 4) and are all separated by the same distance of 3.16 Å. In contrast, the Mo atoms in $LiMoS_2$ occupy two distinct positions in the triclinic unit cell [16] giving rise to short (2.9-3.10 Å) and long (3.44-4.07 Å) Mo-Mo distances. As a result, a chain of diamonds motif of bonded Mo atoms emerges (Fig. 4). This is in qualitative agreement with the structure proposed in an earlier XAFS study [6] and predicted theoretically [5] though the details are significantly different. The "diamond-chain" model is easy to understand using simple electron counting arguments.

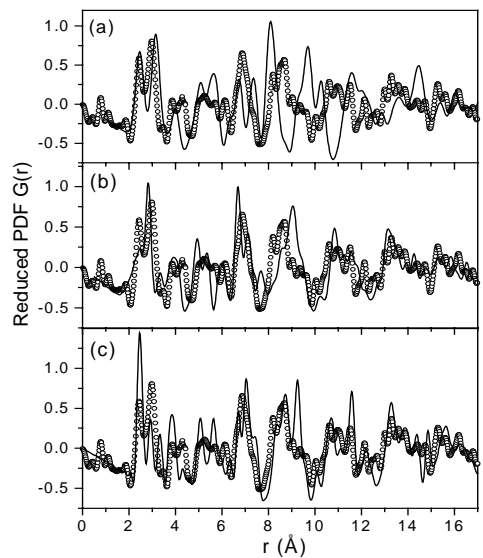

FIG 3. Comparison between the experimental PDF for $LiMoS_2$ (symbols) and model PDFs (full line) for (a) hypothetical hexagonal $LiMoS_2$, (b) hypothetical $1T-LiMoS_2$ (c) triclinic $LiMoS_2$ as predicted in [5].

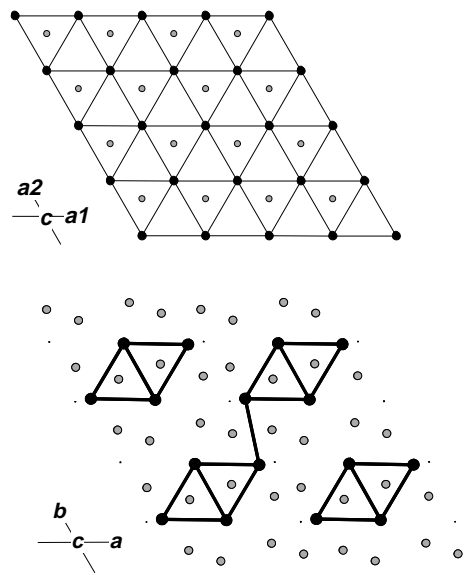

FIG 4. Projection down the *c* axis of the crystal structures of hexagonal $MoS_2$ (up) and triclinic $LiMoS_2$ (down). The large black circles are Mo atoms and the small gray circles are the S atoms. Li atoms are not shown for the sake of clarity.

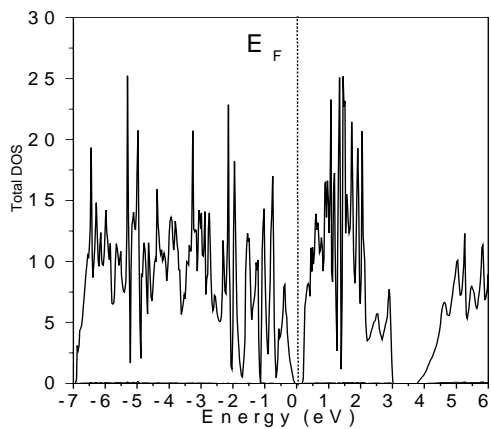

FIG 5. Total Density of States for LiMoS$_2$.

In MoS$_2$ molybdenum is in the 4+ state and has 2 $d$-electrons. It is most stable in a prismatic crystal field resulting in a 1-2-1-1 arrangement of atomic $d$ energy levels. The two electrons both occupy the lowest energy $d_{z^2}$ level which is therefore non-bonding. When Mo gets reduced by the addition of Li it has 3 $d$-electrons. The prismatic coordination is destabilized with respect to octahedral coordination that results in triply degenerate $t_{2g}$ and doubly degenerate $e_g$ levels. One electron goes into each of the 3 $t_{2g}$ levels which point towards neighboring Mo ions. Each Mo can then form bonding interactions with 3 of its neighboring Mo ions resulting in 3 shorter and 3 longer Mo-Mo distances and the diamond-like pattern of distortions shown in Fig. 4.

Using the atomic coordinates determined in this study [16] we calculated the electronic band-structure using the full-potential plane-wave method within density functional theory [17]. As can be seen in Fig. 5 a gap of about 0.2 eV is present in the electronic density of states at the Fermi level. The material should exhibit semiconducting properties as observed [4]. By comparison, in pristine MoS$_2$ the band gap is 1.4 eV.

A recent first-principles calculation predicted a triclinic (S.G. P$\bar{1}$) unit cell for LiMoS$_2$ with Mo atoms arranged in a "diamond-chain" scheme resulting in a 1 eV gap [5]. A comparison between the PDF for the theoretically predicted structure of LiMoS$_2$ and the experimental PDF is shown in Fig. 3(c). The agreement between the two PDF's is satisfactory but the predicted structure does less well than our structural solution (see Fig. 2(a)). Nevertheless, both the present and the previous theoretical studies seem in agreement on the essential structural features of LiMoS$_2$.

In conclusion we have shown that the atomic PDF approach can be confidently employed to solve the structure of nanocrystalline materials. Key to the success of this approach is that both Bragg and the diffuse component of diffraction data are collected over a wide range of wave vectors before converted into the corresponding PDF. Then, like all powder diffraction techniques, the 3-D structure is inferred through modeling. The PDF approach provides a major advantage since testing and refining of structural models can be accomplished even when the diffraction data are very poor in Bragg peaks due to reduced long-range atomic order.

This work was supported by NSF Grant CHE 99-03706, DMR-9817287 and DARPA Grant DAA 695-97-0184. NSLS is supported by the US department of Energy through DE-AC02-98-CH10886.

___